From kni@th.ph.bham.ac.uk Thu Mar 19 12:03 GMT 1998
Date: Thu, 19 Mar 1998 12:02:04 +0000
From: Kirill N Ilinski <kni@th.ph.bham.ac.uk>
MIME-Version: 1.0
To: ass@thsun7.ph.bham.ac.uk
Subject: (no subject)
Content-Transfer-Encoding: 7bit
Content-Disposition: inline; filename="ISbose.tex"

\documentstyle[12pt]{article}
\title{ From Bose condensation to quantum gravity and back}

\author{K.N.Ilinski$^{1,2}$\  
and\ A.S.Stepanenko$^{1,3}$
\\
{\small\it $^{1}$ School of Physics and Space Research, University of
Birmingham,}
\\
{\small\it Birmingham B15 2TT, United Kingdom.} 
\\
{\small\it $^{2}$ Institute of Spectroscopy, Russian Academy of Sciences,} 
\\
{\small\it Troitsk, Moscow region, 142092, Russian Federation.}
\\
{\small\it $^{3}$ Theoretical Department, 
St-Petersburg Nuclear Physics Institute,}
\\
{\small\it Gatchina, St-Petersburg, 188350, Russian Federation.}
}
\date{ }

\begin{document}

\maketitle
\thispagestyle{empty}
\vskip -9cm
\vskip 9cm

\begin{abstract}
We account for the interaction of the Bose-condensed fraction with the normal
phase in an effective dynamical equation such as the Gross-Pitaevskii
equation. We show that the low-energy excitations can be treated as sound
waves with speed dependent on the condensate density. This allows us to
reduce the problem to the calculation of the determinant of the 
Laplace operator on a curved space and apply standard methods of quantum
gravity to get the leading logarithmic contribution of the determinant.
This produces the first quantum correction due to the noncondensed fraction 
to the Gross-Pitaevskii equation for the condensate.
The correction describes an additional
quantum pressure in the condensate and
evaporation-condensation effects. 
\end{abstract}



{\bf 1.} 
Last years there was a new growth of the interest in
Bose-Einstein condensation (in particular, for spatially inhomogeneous
situations) due to the recent success of the experimental
observation of Bose-Einstein condensation for systems of spin polarized
magnetically trapped atomic gases at ultra-low temperatures
\cite{Exp1,Exp2,Exp3} and further investigation of their collective properties
\cite{Exp4}.  These experimental results stimulated the development of a theory
of a nonuniform Bose condensate and its collective excitations.  Moreover, new
exciting problems such as a description of the evolution of the Bose condensate 
from a relaxed trap, stability and dynamics of the collapse of
the condensate for Li$^7$ atoms, collective excitations, 
heating-cooling phenomena, various coherence effects for the condensate 
and so on are being posed both theoretically and experimentally.

All the problems listed above require an 
accurate account of the interaction of the 
condensate with the non-condensed fraction. Formally, the problem can be tackled
using the Bogoliubov-Popov equations which represent a set of coupled linear and 
nonlinear Schr\"odinger equations; the nonlinear Schr\"odinger 
equation (for the condensate fraction) coupled via a potential term 
with a pair of Schr\"odinger equations (for the excited modes) whose
potentials, on their own, depend on the solution of the nonlinear Schr\"odinger
equation. To find an effective equation for the condensate
we need to resolve a pair of Schr\"odinger equations with arbitrary potentials 
and
find the solutions of the equations as functionals of the potentials.
It is obvious that this is hardly a realistic task to perform.

However, as we show in the paper, it is possible to calculate the corrections
from the non-condensed fraction and, surprisingly, they look very simple
in specially chosen notations. The answer comes from exploiting a technique 
originally developed to solve problems in quantum gravity, which seems to be 
quite far from the Bose condensation. To make use of it, we note that the 
low-energy modes of the non-condensed fraction can be considered as sound 
waves whose velocity depends on the density and velocity of the condensate.
This shows that the interaction of the modes with the condensate has the same 
nature as the interaction of matter fields with the background 
gravity field. In this framework the Gross-Pitaevskii  equation is an
analogue of the Einstein equation where the energy-momentum tensor 
is due to presence of quantum matter fields, that is excited modes.
It is not surprising then that to calculate the correction we use
methods from quantum  gravity. More precisely, we use functional integration,
$\zeta$- function renormalization for determinants of elliptic operators and
the Seeley expansion of the heat-kernel for the Laplace operator on a curved 
space 
in an external field. The final form of the correction to the free energy of the 
condensate, written in terms of the curvature tensor then has a very simple 
and manageble form (\ref{Gamma1}).

{\bf 2.}
We are looking for the effective action which describes all physical 
quantities in the system (for an easy introduction to the formalism of 
the effective action see, for example, Appendix A of Ref.\cite{IS1}). 
For example, the effective action provides all of the 
Green's functions in the same approximation used to calculate the 
effective action itself. The equations of motion for the average values of field
operators are given by the equations for a minimum of the effective action.
We obtain them in a hydrodynamic
one-loop approximation. To clarify the
description the hydrodynamical variables, density and velocity, 
should be used instead of microscopic variables.
More precisely, we start with the action 
\begin{equation}
\frac{S}{\hbar} = \int_{-\infty}^{\infty} {\rm d}\tau \int\! {\rm d}V
\left\{ i\psi^{+} \frac{\partial \psi}{\partial \tau}
- \left[\nabla \psi^{+} \nabla \psi +
(\upsilon  - \mu)\psi^{+}\psi + 
 l (\psi^{+}\psi)^2 \right] \right\} \ ,
\label{S1}
\end{equation}
and change field variables to the hydrodynamic ones:
$$
\psi (x,\tau) = \sqrt{\rho(x,\tau)} e^{-i\varphi(x,\tau)}
\qquad , \qquad 
\psi^+ (x,\tau) = \sqrt{\rho(x,\tau)} e^{i\varphi(x,\tau)} \ .
$$
This transforms the action (\ref{S1}) into the form
(up to a complete derivative term):
\begin{equation}
S = \int_{-\infty}^{\infty} {\rm d}\tau \int\! {\rm d}V
\left\{ 
\frac{\partial \varphi}{\partial \tau}\rho -
(\nabla \sqrt{\rho})^2 - \rho (\nabla \varphi)^2 -
(\upsilon - \mu)\rho -
 l \rho^2 \right\} \ .
\label{S2}
\end{equation}
It is not difficult to see that the classical equations of motion for the
action lead to the equations:
\begin{equation}
\frac{\partial {\bf c}}{\partial \tau} + \nabla \left(
\frac{{\bf c}^2}{2}
+2v  - 2\mu +
4 l \rho - \frac{2}{\sqrt{\rho}}\nabla^2 \sqrt{\rho}\right)  = 0 \ ,
\label{eq3}
\end{equation}
\begin{equation}
\frac{\partial \rho}{\partial \tau} +
 \nabla ({\bf c}  \cdot\rho)
  = 0 \ ,
\label{eq4}
\end{equation}
which are equivalent to the Gross-Pitaevskii equation \cite{Stringari}
after the introduction of the velocity variable ${\bf c}=-2\nabla \varphi$
instead of $\varphi$. These equations look like
hydrodynamic equations for an irrotational compressible fluid which is 
the condensate superfluid.

Now let us shift our variables in (\ref{S1}) by the zero-order mean 
field solution (the condensate amplitude) to find the one loop quantum 
correction for the effective action:
$$
\psi \rightarrow \psi + \chi \qquad , \qquad 
\psi^{+} \rightarrow \psi^{+} + \chi^{+}
$$
and keep only quadratic terms of $\chi$ and $\chi^{+}$ 
since it can be shown~\cite{IS1} that this is the only 
relevant contribution for the case of trapped gases. Then the action 
(\ref{S1}) transforms to 
$S(\psi,\psi^{+}) + S_1(\psi,\psi^{+},\chi,\chi^{+})$ where
the term $S_1$ can be written as:
\begin{eqnarray}
S_1(\psi,\psi^{+},\chi,\chi^{+}) &=& 
\int_{-\infty}^{\infty} {\rm d}\tau \int\! {\rm d}V
\Biggl\{ i\chi^{+} \frac{\partial \chi}{\partial \tau}
- \Bigl[\nabla \chi^{+} \nabla \chi +
\{\dot{\varphi} - (\nabla\varphi)^2 + 2l\rho \}\chi^{+}\chi
\nonumber\\
 &&\quad - l {\rm e}^{2i\varphi}\rho \chi^{+} \chi^{+}
 - l {\rm e}^{-2i\varphi}\rho \chi \chi 
\Bigl] \Biggl\} \ .
\nonumber
\end{eqnarray}
In the last equation the potential $\upsilon$
was excluded using the mean field equation of motion neglecting 
the kinetic term. The Hamiltonian corresponding to $S_1$ is
$$
H = \int\! {\rm d}V
\left[\nabla \hat\chi^{+} \nabla \hat\chi +
\{ \dot{\varphi} - (\nabla\varphi)^2 + 2l\rho \} \hat\chi^{+}\hat\chi 
 - l {\rm e}^{2i\varphi}\rho \hat\chi^{+} \hat\chi^{+}
 - l {\rm e}^{-2i\varphi}\rho \hat\chi \hat\chi 
\right] \ .
$$
Now we perform the following canonical transformation:
$$
\hat\chi = 
\sqrt{\rho}{\rm e}^{i\varphi}\left[\frac{\hat\sigma}{2\rho} + 
i\hat\alpha\right]
\qquad , \qquad
\hat\chi^{+} = 
\sqrt{\rho}{\rm e}^{-i\varphi}\left[\frac{\hat\sigma}{2\rho} - 
i\hat\alpha\right] \ .
$$
The Hamiltonian takes the form
\begin{equation}
\begin{array}{l}
H = \int\! {\rm d}V
\Biggl[
   \rho (\nabla\hat\alpha)^2
 + \rho\dot\varphi \hat\alpha^2
 + \{ l + \frac{\displaystyle\dot\varphi}{\displaystyle 4\rho} \} \hat\sigma^2
 + \hat\sigma (\nabla\varphi\nabla\hat\alpha)
 - \hat\alpha (\nabla\varphi\nabla\hat\sigma)
 + \frac{1}{2}\Delta\varphi\hat\sigma\hat\alpha 
\nonumber\\ 
-\frac{1}{\displaystyle 4\rho^2}\hat\sigma\nabla\rho\nabla\hat\sigma 
 + \frac{\displaystyle (\nabla\rho)^2}{\displaystyle 4\rho}\hat\alpha^2
 + \frac{1}{\displaystyle 4\rho}(\nabla\hat\sigma)^2
 + \frac{\displaystyle \nabla\rho\nabla\varphi}{\displaystyle \rho}
	\hat\alpha\hat\sigma
 + \frac{\displaystyle (\nabla\rho)^2}{\displaystyle 16\rho^3}\hat\sigma^2
 + \hat\alpha\nabla\rho\nabla\hat\alpha
\Biggr] \ .
\end{array}
\nonumber
\end{equation}
Last eight terms  vanish in the hydrodynamical limit since
$\rho$ is a large variable ($\nabla\varphi\sim\sqrt{l\rho}$ 
and $\dot\varphi\sim l\rho$), so we have
$$
H = \int\! {\rm d}V
\Biggl[
   \rho (\nabla\hat\alpha)^2
 + \rho\dot\varphi \hat\alpha^2
 + \{ l + \frac{\displaystyle\dot\varphi}{\displaystyle 4\rho} \} \hat\sigma^2
 + \hat\sigma (\nabla\varphi\nabla\hat\alpha)
 - \hat\alpha (\nabla\varphi\nabla\hat\sigma)
\Biggr] \ .
$$
The corresponding action (in the Feynman functional integral) is
$$
S_1 = 
\int_{-\infty}^{\infty} {\rm d}\tau \int\! {\rm d}V
\Biggl\{ \frac{\partial \alpha}{\partial \tau}\sigma
 - \Bigl[
   \rho (\nabla\alpha)^2
 + \rho\dot\varphi \alpha^2
 + \{ l + \frac{\displaystyle\dot\varphi}{\displaystyle 4\rho} \} \sigma^2
 + 2\sigma (\nabla\varphi\nabla\alpha)
\Bigr]
\Biggr\} \ .
$$
Integrating the functional integral over the  field $\sigma$ we get  
the action for 
$\alpha$ field only:
$$
S_1 = \int_{-\infty}^{\infty} {\rm d}\tau \int\! {\rm d}V
\left\{ -\rho (\nabla \alpha)^2 +
f\Bigl(\frac{\partial\alpha}{\partial \tau} 
 - 2 \nabla \varphi \nabla \alpha 
\Bigr)^2
 - \rho\dot\varphi\alpha^2 
\right\} \ ,
$$
where
$
f = \frac{1}{4l + \dot\varphi/\rho}\ 
$. The corresponding determinant from the integration does not 
contribute to the effective action at this order of the perturbation theory.

Now we will once again use the existence of the large  parameter $\rho$.
Let us introduce the following large number 
$\rho_0 \equiv {\rm max}\, \{\rho \}$ such that
$\tilde{\rho} = \frac{\rho}{\rho_0} \sim 1$, and new space-time variables 
$t \equiv 4l\rho_{0}\tau $ and $y_i\equiv \sqrt{ 4l\rho_{0}} x_i$. The action 
$S_1$ can be represented as
\begin{eqnarray}
S_1\!\!& =&\!\! 
\frac{1}{\sqrt{64\rho_0 l^3}} 
\int_{-\infty}^{\infty} {\rm d}t \int\! {\rm d}\tilde V\
\left\{ -\tilde{\rho} (\tilde\nabla \alpha)^2 +
\tilde f\Bigl(
\frac{\partial\alpha}{\partial t}
 - \tilde{\bf v} \tilde\nabla \alpha\Bigr)^2
 - \tilde\rho\dot\varphi\alpha^2 
\right\} 
\nonumber\\
\!\!& \equiv&\!\!
 - \frac{1}{\sqrt{64\rho_0 l^3}}
\int_{-\infty}^{\infty} {\rm d}t \int\! {\rm d}\tilde V\
\left[
A^{\mu \nu} \partial_{\mu}\alpha \partial_{\nu}\alpha
 + \tilde\rho\dot\varphi\alpha^2
\right]\ ,
\nonumber
\end{eqnarray}
where
$$
\tilde{\bf v} = 2\tilde\nabla \varphi\ , \quad
\tilde f = 4lf = \frac{1}{1 + \dot\varphi/\tilde\rho}
$$
and the matrix $A$ has the form
$$
A = 
\left(  
\begin{array}{cccc}
-\tilde f & \tilde f\tilde{v}_1  & \tilde f\tilde{v}_2  &\tilde f \tilde{v}_3 \\
\tilde f\tilde{v}_1 & \tilde{\rho}-\tilde f\tilde{v}_1^2
	 & -\tilde f\tilde{v}_1\tilde{v}_2 &  -\tilde f\tilde{v}_1\tilde{v}_3 \\
\tilde f\tilde{v}_2 & -\tilde f\tilde{v}_1\tilde{v}_2 
	 & \tilde{\rho}-\tilde f\tilde{v}_2^2 &  -\tilde f\tilde{v}_2\tilde{v}_3 
\\
\tilde f\tilde{v}_3 & -\tilde f\tilde{v}_1\tilde{v}_3 
 	 & -\tilde f\tilde{v}_2\tilde{v}_3 & \tilde{\rho}-\tilde f\tilde{v}_3^2  
\\
\end{array} 
\right) \ .
$$
Our next step is to cast the action in the covariant form. To do this we 
introduce auxiliary metric $\tilde{g}_{\mu\nu}$ such that 
$$
A^{\mu \nu} = \frac{\tilde{g}^{\mu\nu}}{\sqrt{-{\rm det}\, 
(\| \tilde{g}^{\mu\nu} \|)}} \ , \qquad
\tilde{g}^{\mu \nu} = \frac{A^{\mu\nu}}{\sqrt{-{\rm det}\, (\| A^{\mu\nu} \|)}} 
\ .
$$
This leads to the form for the metric $\|\tilde{g}_{\mu\nu}\|$:
$$
\|\tilde{g}_{\mu\nu}\|= \tilde{\rho}^{1/2}\tilde{f}^{1/2}
\left(  
\begin{array}{cccc}
-\tilde{\rho}/\tilde f + \tilde{v}^2 & \tilde{v}_1 & \tilde{v}_2 & \tilde{v}_3 
\\
\tilde{v}_1 & 1 & 0  &  0 \\
\tilde{v}_2 & 0 &  1 &  0  \\
\tilde{v}_3 & 0 & 0 &  1  \\
\end{array} \right) \ .
$$
In this metric the action has a covariant form
$$
S_1 =
 -\frac{1}{\sqrt{64\rho_0 l^3}}
\int\! {\rm d}y\ \sqrt{-\tilde{g}}
 \left[\alpha  \tilde{\Delta} \alpha + \tilde E \alpha^2\right] 
$$
where
$$
\tilde E = \frac{\tilde \rho\dot\varphi}{\sqrt{-\tilde g}}
 = \frac{\dot\varphi}{\sqrt{\tilde\rho\tilde f}}\ ,
 \qquad
 \tilde{\displaystyle\Delta} 
= -\frac{1}{\sqrt{-\tilde{g}}} \partial_{\mu} 
\tilde{g}^{\mu \nu}  \sqrt{-\tilde{g}} \partial_{\nu} \ .
$$
Now the effective action is 
$\Gamma _{1} = - \frac{1}{2}{\rm tr}\, \ln [(64\rho_0 l^3)^{-1/2} 
(\tilde{\displaystyle\Delta}+\tilde E)]$
with Laplace operator 
$\tilde{\displaystyle\Delta} 
= -\frac{1}{\sqrt{-\tilde{g}}} \partial_{\mu} 
\tilde{g}^{\mu \nu}  \sqrt{-\tilde{g}} \partial_{\nu} $.

At this point the quantum gravity analogy comes into the game and let us
evaluate $\Gamma_1$ using $\zeta$-function regularization~\cite{Schwarz}
for $\frac{1}{2}{\rm tr}\, \ln [(64\rho_0 l^3)^{-1/2} 
(\tilde{\displaystyle\Delta}+\tilde E)]$ after performing the Wick rotation to
make the operator elliptic. In this regularization
the quantity is finite and has the following property:
$$
{\rm tr}\, \ln [(64\rho_0 l^3)^{-1/2} 
(\tilde{\displaystyle\Delta}+\tilde E)]
 = {\rm tr}\, \ln [ 
\tilde{\displaystyle\Delta}+\tilde E ]
 - \frac{1}{2} {\rm tr}\, \ln (64\rho_0 l^3) 
\Bigl(\Phi_0 (\tilde{\displaystyle\Delta}+\tilde E )
 - L(\tilde{\displaystyle\Delta}+\tilde E) \Bigr) \ ,
$$
where 
$\Phi_0(\tilde{\displaystyle\Delta}+\tilde E)\equiv
\int\!  {\rm d}y{\rm d}\tau\ 
\sqrt{-\tilde g} \Psi_0(\tilde{\displaystyle\Delta}+\tilde E)$
is so-called zeroth Seeley coefficient and $L$ is a 
number of zero modes. For our goal is important to realize that
in the regularization 
${\rm tr}\, \ln [ \tilde{\displaystyle\Delta}+\tilde E]$ 
is of order of the zeroth Seeley coefficient 
 which, on  its own, is of order unity.

The existence of the large parameter in the system allows us to make 
the next step.
Indeed, the multiplier $(64\rho_0 l^3)^{-1/2}$ gives us the possibility to
express the main contributions to the determinant such as
$$
{\rm tr}\, \ln [(64\rho_0 l^3)^{-1/2} 
(\tilde{\displaystyle\Delta}+\tilde E)]
\sim 
 - \frac{1}{2} {\rm tr}\, \ln (64\rho_0 l^3) \int dxdt
\Bigl(\Psi_0 (\tilde{\displaystyle\Delta}+\tilde E )
 - L(\tilde{\displaystyle\Delta}+\tilde E) \Bigr)
$$
The Seeley coefficient $\Psi_{0}$  for the Laplace operator in 4D
is well-known and can be found, for example, in Ref.~\cite{Schwarz}:
$$
\Psi_{0} = \frac{\sqrt{-g}}{(4\pi)^2} ( -\frac{1}{30}\nabla^2 R 
+ \frac{1}{72}R^2 -
\frac{1}{180}R_{\mu\nu} R^{\mu\nu}
 + \frac{1}{180}R_{\mu\nu\sigma\rho} R^{\mu\nu\sigma\rho} 
$$
$$
+\frac{1}{6}ER  
+ \frac{1}{2}E^2 -\frac{1}{6}\nabla ^2 E ) \ .
$$
where $R$, $R_{\mu\nu}$ and $R_{\mu\nu\sigma\rho}$ 
are the scalar curvature,  Ricci and Riemann tensors 
correspondingly~\cite{DNFW}.

In all physical applications the number of zero-modes does not change
and we can drop  the term with 
$L(\tilde{\displaystyle\Delta}+\tilde E)$.
Returning to the initial variables and performing inverse Wick rotation, 
all curvature tensors  and the metric
are written in the ``physical" variables (i.e. without tildes).
Moreover $\ln (64\rho_0 l^3) \gg \ln (\tilde{\rho})$ so that we can substitute
$\ln (64\rho l^3)$ instead of $\ln (64\rho_0 l^3)$. Summarizing, 
we obtain an expression for the first quantum correction to the
effective action:
\begin{equation}
\Gamma _{1} = \frac{1}{4} \int\!  {\rm d}x{\rm d}\tau\ \sqrt{-g}
\ln (64\rho l^3) \Psi_0(\Delta +E) \ ,
\label{Gamma1}
\end{equation}
with the metric 
$$
\|g_{\mu\nu}\|= \rho^{1/2}f^{1/2}
\left(  
\begin{array}{cccc}
-\rho/f + v^2 &  v_1  &  v_2  &  v_3  \\
v_1 & 1 & 0  &  0 \\
v_2 & 0 &  1 &  0  \\
v_3 & 0 & 0 &  1  \\
\end{array} \right) \ .
$$
and 
$$
v_i\equiv 2\partial_i\varphi\ ,\quad 
f = \frac{1}{4l + \dot\varphi/\rho}\ ,\quad
E = \frac{\dot\varphi}{\sqrt{\rho f}}\ .
$$
Equation (\ref{Gamma1}) is the central equation of the paper.

Miminizing the effective action (\ref{S2}) with the correction (\ref{Gamma1}), 
the equations of motion for the condensate can be derived as:
$$
\frac{\partial \varphi}{\partial \tau}
 - (\nabla \varphi)^2
 - v  + \mu
 - 2 l \rho
 + \frac{1}{4} \ln (64\rho l^3)
\Biggl[
\frac{\partial g_{\mu\nu}}{\partial \rho}
\frac{\delta \Psi_{0}}{\delta g_{\mu\nu}}
 + \frac{\partial E}{\partial \rho} 
\frac{\delta \Psi_{0}}{\delta E}
\Biggr]
 = 0 \ ,
$$
$$
 - \frac{\partial \rho}{\partial \tau}
 + 2 \nabla (\nabla\varphi  \cdot\rho)
 + \frac{1}{4} \ln (64\rho l^3)
\Biggl[
\int\!  {\rm d}^4 y\ 
\frac{\delta g_{\mu\nu}(x)}{\delta \varphi(y)}
\frac{\delta \Psi_{0}}{\delta g_{\mu\nu}(x)}
 - \frac{\partial}{\partial\tau}
\left(
\frac{\partial E}{\partial \dot\varphi} 
\frac{\delta \Psi_{0}}{\delta E}
\right)
\Biggr] 
= 0 \ .
$$
The equations substitute for the Gross-Pitaevskii equation and generalize Eqns 
(\ref{eq3},\ref{eq4}) to account for the
interaction of the Bose-condensate with the non-condensed fraction 
in the one-loop approximation.

{\bf 3.} In conclusion, we have calculated the one-loop quantum correction 
for the effective action of the condensate. The correction comes from
the interaction of the condensate with low-energy excited modes. 
To this end, we considered the excited modes as quantum particles moving in
curved space whose metric (gravity field) was defined by the density and the 
velocity of the condensate. The equations of motion for the condensate are then
analogous to the equations for the gravitational field where the right hand side 
(the energy-momentum tensor) is due to the presence of matter fields. 
The analysis does not depend on details of the external 
potential and can be used for variety of problems. The only simplification
is the common hydrodynamic approximation.

\vspace{0.5cm}

\noindent
{\it Acknowledgment. }
We are grateful to Keith Burnett and Mike Gunn for 
the discussions of the problem.  
This work was supported by the Grant 
of Russian Fund of Fundamental Investigations 
N 95-01-00548 and by the UK EPSRC Grants GR/L29156, GR/K68356.


\begin{thebibliography}{99} 
\bibitem{Exp1}
M.~H. Anderson, J.~R. Ensher, M.~R. Matthews, C.~E. Wieman, and E.~A. Cornell,
Science {\bf 269}, 198 (1995);

\bibitem{Exp2}
C.~C. Bradley, C.~A. Sackett, J.~J. Tollett, and R.~G. Hulet, 
Phys.~Rev.~Lett. {\bf 75}, 1687 (1995);

\bibitem{Exp3} K.~B. Davis, M.-O. Mewes, M.~R. Andrews, N.~J. van Druten, 
D.~S. Durfee, D.~M. Kurn, and W. Ketterle, Phys.~Rev.~Lett. {\bf 75}, 3969 
(1995);

\bibitem{Exp4} D.~S. Jin, J.~R. Ensher, M.~R. Matthews, C.~E. Wieman and
E.~A. Cornell, Phys.~Rev.~Lett.{\bf 77}, 420 (1996);

\bibitem{IS1} K.N.Ilinski and A.S.Stepanenko: {\it 
 First quantum corrections for a hydrodynamics of a nonideal Bose gas}, 
preprint cond-mat/9607202; can be found 
http://xxx.lanl.gov/abs/cond-mat/9607202;

\bibitem{Stringari} S.Stringari, Phys.Rev.Lett. {\bf 77}, 2360 (1996);

\bibitem{Schwarz} Albert S. Schwarz: {\it Quantum Field Theory and Topology},
Springer-Verlag, 1993;

\bibitem{DNFW} B.A.Dubrovin, A.T.Fomenko, S.P.Novikov:
{\it Modern Geometry -- Methods and Applications}, Springer-Verlag, 1984;
S.Weinberg: {\it Gravitation and cosmology: principles and applications of 
the general theory of relativity}, John Wiley \& Sons,  
Inc. New York London Sydney Toronto, 1972.

\end{thebibliography}
\end{document}